\documentclass[11pt]{amsart}
\usepackage[square, authoryear]{natbib}
\usepackage{amsmath,amssymb,amsthm,latexsym,euscript}

\newcommand{\extd}{\ensuremath{\mathbf{d}}}
\newcommand{\lie}[2]{\ensuremath{{\pounds}_{#1}{#2}}}
\newcommand{\intp}[2]{\ensuremath{{\mathbf{i}}_{#1}{#2}}}

\theoremstyle{plain} \newtheorem{Prop}{Proposition}
\theoremstyle{plain} \newtheorem{Lem}{Lemma}

\begin{document}

\begin{abstract}
Here we carry out computations that help clarify the Lagrangian and Hamiltonian
structure of compressible flow.  The intent is to be pedagogical and rigorous, providing concrete
examples of the theory outlined in \cite{HoMaRa1998} and \cite{MaRaWe1984}.
\end{abstract}

\title[User's Guide to Semidirect Product Reduction Theory]{Semidirect Product Reduction Theory: \\A User's Guide}
\author{H.S. Bhat}
\thanks{Computations from sections 1-3 carried out with R. Fetecau and M. West.}
\address{Control and Dynamical Systems 107-81,
  California Institute of Technology,
  Pasadena CA 91125, USA}
\email{bhat@cds.caltech.edu}
\date{\today}
\maketitle

\tableofcontents

\section{Spaces and Actions}
We start with the group $G = \text{Diff}(M)$ and the vector space 
$V = \EuScript{C}^{\infty}(M)$.  Multiplication in the group $G$ is denoted
by a single centered dot, i.e. given $\gamma_1, \gamma_2 \in G$,
$$
\gamma_1 \cdot \gamma_2 = \gamma_1 \circ \gamma_2.
$$
For our purposes, we write the dual of $V$ as
$V^{\ast} = \text{Den}(M)$, the space of all $\EuScript{C}^{\infty}$ volume
forms on $M$. If the manifold $M$ has dimension $n$, then
$$
\text{Den}(M) = \Omega^{n}(M),
$$
the space of smooth $n$-forms on $M$.
We pair an element $f \in V$ with $\mu \in V^{\ast}$ by
$$
\langle \mu, f  \rangle = \int_{M} f \, \mu.
$$

\subsection{$G$ action on $V$} Let's define an action of $G$ on $V$ by
$$
\Phi^{V}_{\gamma} (f) = \gamma^{\ast}(f).
$$
What we have here is a map which associates a linear isomorphism $\Phi^V_{\gamma}$ to each element $\gamma \in G$.
We will sometimes write this action more compactly using concatenation:
$$
\gamma f = \Phi^{V}_{\gamma} (f) = \gamma^{\ast}(f).
$$
The action is clearly a right action:
\begin{align*}
(\gamma_1 \cdot \gamma_2) f &= (\gamma_1 \circ \gamma_2)^{\ast}(f) \\
 &= {\gamma_2}^{\ast} ({\gamma_1}^{\ast} (f)) \\
 &= \gamma_2 (\gamma_1 f)
\end{align*}

\subsection{Induced $G$ Action on $V^\ast$}
The $G$ action on $V$ induces a right action on $V^\ast$ via the inverse of
the dual of $\Phi_{\gamma}^V$.  We now calculate the dual
isomorphism $\left( \Phi_{\gamma}^{V} \right)^{\ast}$.
Given $f \in V$ and $\omega \in V^{\ast}$,
\begin{align*}
\langle \left( \Phi_{\gamma}^{V} \right)^{\ast} (\omega), f \rangle 
&= \langle \omega, \Phi_{\gamma}(f) \rangle \\
&= \langle \omega, \gamma^{\ast} (f) \rangle \\
&= \int_{M} \gamma^{\ast} (f) \, \omega \\
&= \int_{M} \gamma^{\ast} (f \, \gamma_{\ast}(\omega)) \\
&= \int_{M} f \, \gamma_{\ast}(\omega) \\
&= \langle \gamma_{\ast}(\omega), f \rangle.
\end{align*}
This shows that
$$
\left( \Phi_{\gamma}^{V} \right)^{\ast} (\omega) = \gamma_{\ast}(\omega)
$$
Now we can define the action of $G$ on $V^{\ast}$:
\begin{align*}
\Phi_{\gamma}^{V^{\ast}}(\omega) &= 
\left[ \left( \Phi_{\gamma}^{V} \right)^{\ast} \right]^{-1} (\omega) \\
 &= (\gamma_{\ast})^{-1}(\omega) \\
 &= \gamma^{\ast}(\omega)
\end{align*}
The proof that this is a right action of $G$ on $V^{\ast}$ is identical to 
the case of $G$ acting on $V$.

\subsection{Semidirect Product Group and Algebra} 
We will be concerned with the semidirect 
product space $G \circledS V^{\ast}$, which is a group consisting of the 
set $G \times V^{\ast}$ equipped with the product
$$
(\gamma_1, \omega_1) \cdot (\gamma_2, \omega_2) = 
(\gamma_1 \cdot \gamma_2, \omega_2 + \gamma_2 \omega_1).
$$
The Lie algebra $\mathfrak{g}$ of $G$ consists of the space $T_{e} G$ 
equipped with the appropriate bracket.  As can be shown without much 
difficulty, for a given diffeomorphism $\gamma \in G$,
$$
T_{\gamma} G = 
\{ \text{ all maps } X: M \to TM \text{ such that } \pi_{M} \circ X = \gamma \}.
$$
Therefore,
$$
T_{e} G = \mathfrak{X}(M) = \{ \text{ all vector fields } X : M \to TM \},
$$
since $\pi_{M} \circ X = \mathrm{id}_M$ is precisely the condition that makes 
$X$ a vector field.  (Clearly $e \in G$ is the identity map $\mathrm{id}_M$ on
$M$.)

\begin{Prop}
The bracket $[\xi, \eta]_{\mathfrak{g}}$ on the Lie algebra $\mathfrak{g}$ is
\emph{minus} the Jacobi-Lie bracket of vector fields on $M$.  In symbols,
$$
[\xi,\eta]_{\mathfrak{g}} = - [\xi,\eta]_M.
$$
\end{Prop}
Proof: Let $X^R_{\zeta}$ denote the \emph{right}-invariant 
vector field associated with $\zeta$, i.e. 
$X^R_{\zeta} (\gamma) = \zeta \circ \gamma$.  Keep in mind that
$X^R_{\zeta}$ is a vector field on $\text{Diff}(M)$.

Let's verify that $X^R_{\zeta}$ is right-invariant.
Set $R_{\gamma} (\beta) = \beta \circ \gamma$, 
right-translation on $\text{Diff}(M)$.
Then, for all maps $\zeta_{\alpha} : M \to TM$ such that 
$\pi_{M} \circ \zeta = \alpha$,
$$
T_{\alpha} R_{\gamma} \cdot \zeta_{\alpha} = \zeta \circ \gamma,
$$
It is now clear that
\begin{align*}
T_{e} R_{\gamma} \cdot X^R_{\zeta} (\eta) &= (\zeta \circ \eta) \circ \gamma \\
 &= \zeta \circ \eta \circ \gamma = \zeta \circ R_{\gamma} (\eta) \\
 &= X^R_{\zeta} \circ R_{\gamma} (\eta),
\end{align*}
as required.  With this in mind, we use the local formula for the bracket
(again, the Jacobi-Lie bracket of vector fields on $\text{Diff}(M)$):
\begin{align*}
[X^R_{\xi}, X^R_{\eta}](e) &= 
DX^R_{\xi}(e) \cdot X^R_{\eta}(e) - DX^R_{\eta}(e) \cdot X^R_{\xi}(e) \\
 &= D(\xi) \cdot \eta - D(\eta) \cdot \xi
\end{align*}
and this is the local formula for the Jacobi-Lie bracket of vector fields on $M$.
We have therefore shown that
\begin{equation}
\label{eqn:brakfirst}
[X^R_{\xi}, X^R_{\eta}](e) = [\xi, \eta]_{M}.
\end{equation}
Now we are almost done.  Typically, we define a bracket on $\mathfrak{g}$ via
\begin{equation}
\label{eqn:brakdef}
[\xi,\eta]_{\mathfrak{g}} := [X^L_{\xi},X^L_{\eta}](e)
\end{equation}
where $X^L_{\xi}$ is the \emph{left}-invariant vector field associated with
$\xi \in \mathfrak{g}$.
\begin{Lem}
Defining a bracket on the Lie algebra $\mathfrak{g}$ as in (\ref{eqn:brakdef})
implies that
\begin{equation}
\label{eqn:rightbrak}
[X^R_{\xi},X^R_{\eta}](e) = - [\xi,\eta]_{\mathfrak{g}}.
\end{equation}
\end{Lem}
Proof: Introduce two spaces:
\begin{align*}
\mathfrak{X}^R &= \text{Lie algebra of all right-invariant vector fields on Diff}(M) \\
\mathfrak{X}^L &= \text{Lie algebra of all left-invariant vector fields on Diff}(M)
\end{align*}
Start with the diffeomorphism $\phi$ on $G$ given by $\phi(g) = g^{-1}$.
One can show that for $X$ a left-invariant vector field,
$f(X) = \phi_{\ast}(X)$ is a right-invariant vector field.  Indeed the map
$f$ defined in this way is a Lie algebra isomorphism from $\mathfrak{X}^R$
to $\mathfrak{X}^L$.  Furthermore one can compute
$$
T_{e} \phi \cdot \xi = -\xi
$$
for all $\xi \in \mathfrak{g}$.  Now we have, for $\eta \in \mathfrak{g}$,
\begin{align*}
f(X^L_{\eta})(e) &= (T \phi \circ X^L_{\eta} \circ \phi^{-1})(e) \\
 &= T_{e} \phi \circ X^L_{\eta}(e) \\
 &= T_{e} \phi \circ (T(\text{id}_G) \circ \eta) \\
 &= T_{e} \phi \cdot \eta \\
 &= -\eta.
\end{align*}
Then
\begin{align*}
[X^R_{\xi}, X^R_{\eta}](e) &= [\xi, \eta]_{M} \\
 &= [f(X^L_{-\xi}), f(X^L_{-\eta})]_{M} \\
 &= [\phi_{\ast}(X^L_{-\xi}), \phi_{\ast}(X^L_{-\eta})]_{M} \\
 &= \phi_{\ast} [X^L_{-\xi}, X^L_{-\eta}](e) \\
 &= f(X^L_{[-\xi, -\eta]_{\mathfrak{g}}})(e) \\
 &= -[-\xi, -\eta]_{\mathfrak{g}} \\
 &= -[\xi, \eta]_{\mathfrak{g}},
\end{align*}
which is what was required.  Putting together (\ref{eqn:rightbrak}) with (\ref{eqn:brakfirst}) finishes the proof of the proposition.

\subsection{The dual algebra $\mathfrak{g}^\ast$} We will
take $\mathfrak{g}^{\ast}$ to be the space of one-form densities:
$$
\mathfrak{g}^{\ast} = \{ \omega \otimes \mu : \omega \in \Lambda^1(M), \mu \in \text{Den}(M)\}
$$
Then for $\xi \in \mathfrak{g}$, $\theta \otimes \mu \in \mathfrak{g}^{\ast}$ 
the pairing between the two elements is given by
$$
\langle \theta \otimes \mu, \xi \rangle = \int_{M} \theta(\xi) \, \mu.
$$

\subsection{Induced $\mathfrak{g}$ action on $V$} The action of $G$ on $V$ induces an action of $\mathfrak{g}$ on $V$.  We use
$\Phi : G \to GL(V,V)$ to denote the original action:
$\Phi(\gamma) = \gamma^{\ast}( \, \cdot \, )$.  Then we can compute
$T_{e} \Phi : \mathfrak{g} \to L(V,V)$ as follows.  Fix $\xi \in \mathfrak{g}$
and write $\xi$ as a tangent vector to a curve 
$\gamma^{\epsilon}$ on $G=\text{Diff}(M)$:
$$
\xi = \left. \frac{d}{d \epsilon} \right|_{\epsilon = 0} \gamma^{\epsilon}
 \text{ where } \gamma^0 = \text{id}_G.
$$
Then for $f \in V$,
$$
(T_{e} \Phi \cdot \xi)(f) = \left. \frac{d}{d \epsilon} \right|_{\epsilon=0}
 (\gamma^{\epsilon})^{\ast}(f) = \lie{\xi}{f},
$$
by the dynamic definition of the Lie derivative.  From now on we will use
concatenation to denote the action of $\mathfrak{g}$ on $V$, so that for
$\xi \in \mathfrak{g}$ and $f \in V$,
$$
\xi f = \lie{\xi}{f}.
$$

\subsection{Induced $\mathfrak{g}$ action on $V^{\ast}$} 
The $\mathfrak{g}$-action on $V^{\ast}$ is defined as \emph{minus} 
the dual map
(i.e. linear algebraic adjoint) of the $\mathfrak{g}$-action on $V$.  We can also
start with the action of $G$ on $V^{\ast}$; this action induces a
$\mathfrak{g}$-action on $V^{\ast}$.  Both $\mathfrak{g}$-actions on $V^{\ast}$
are in fact the same, as we will now show.  Take $\mu \in V^{\ast}$, $f \in V$,
and $\xi \in \mathfrak{g}$.  Then
\begin{align*}
\langle \mu, \lie{\xi}{f} \rangle &= 
\int_{M} \left( \lie{\xi}{f} \right) \mu \\
&= \int_{M} \left( \lie{\xi}{(f \mu)} - f \lie{\xi}{\mu} \right) \\
&= \int_{M} \left( \extd \intp{\xi}{(f \mu)} + \intp{\xi}{\extd (f \mu)} \right) -
   \int_{M} f \lie{\xi}{\mu} 
\end{align*}
Since $f \mu$ is an $n$-form, $d(f \mu) = 0$.  Also, since $\partial M = \emptyset$, Stokes' theorem implies that
$$
\int_{M} \extd \intp{\xi}{(f \mu)} = 0.
$$
Therefore, we have
$$
\langle \mu, \lie{\xi}{f} \rangle = \langle -\lie{\xi}{\mu}, f \rangle,
$$
so the $\mathfrak{g}$-action on $V^{\ast}$ is
$$
\xi \mu = \lie{\xi}{\mu}.
$$
Verifying the calculation using the alternate approach, we start with the
$G$-action on $V^{\ast}$ denoted by $\Psi : G \to GL(V^{\ast},V^{\ast})$,
where for each $\gamma \in G$, $\Psi(\gamma) = \gamma^{\ast}(\, \cdot \,)$.
Then we compute $T_{e} \Psi : \mathfrak{g} \to L(V^{\ast},V^{\ast})$.  As
before, we fix $\xi \in \mathfrak{g}$ and write
$$
\xi = \left. \frac{d}{d \epsilon} \right|_{\epsilon = 0} \gamma^{\epsilon}
\text{ where } \gamma^{0} = \text{id}_G.
$$ 
Then
$$
(T_{e} \Psi \cdot \xi) (\mu) = 
\left. \frac{d}{d \epsilon} \right|_{\epsilon = 0}
(\gamma^{\epsilon})^{\ast}(\mu) = \lie{\xi}{\mu}
$$
by the dynamic definition of the Lie derivative.

\subsection{Adjoint action (group)}
We start with the conjugation map
$$
I_{\gamma} (\eta) = \gamma \cdot \eta \cdot \gamma^{-1}.
$$
We can then write the adjoint action as
\begin{align*}
\text{Ad}_{\gamma} \xi &= T_{e} I_{\gamma} \cdot \xi \\
 &= T\gamma \circ \xi \circ \gamma^{-1} \\
 &= \gamma_{\ast} \xi
\end{align*}

\subsection{Dual of adjoint action (group)} 
Next on our list is the $\text{Ad}^{\ast}$ action.  We take
$\omega \otimes \mu \in \mathfrak{g}^{\ast}$, $\xi \in \mathfrak{g}$,
and $\gamma \in G$.  Then the calculation runs as follows:
\begin{align*}
\langle \text{Ad}^{\ast}_{\gamma} \omega \otimes \mu, \xi \rangle &=
\langle \omega \otimes \mu, \text{Ad}_{\gamma} \xi \rangle \\
&= \langle \omega \otimes \mu, \gamma_{\ast} \xi \rangle \\
&= \int_{M} \intp{\gamma_{\ast} \xi}{\omega} \, \mu \\
&= \int_{M} \gamma_{\ast} (\intp{\xi}{\gamma^{\ast} \omega}) \, \mu \\
&= \int_{M} \gamma_{\ast} (\intp{\xi}{\gamma^{\ast} \omega} \, \gamma^{\ast} \mu) \\
&= \int_{M} \intp{\xi}{\gamma^{\ast} \omega} \, \gamma^{\ast} \mu \\
&= \langle \gamma^{\ast} \omega \otimes \gamma^{\ast} \mu, \xi \rangle \\
&= \langle \gamma^{\ast} (\omega \otimes \mu), \xi \rangle,
\end{align*}
so
$$
\text{Ad}^{\ast}_{\gamma} (\omega \otimes \mu) = \gamma^{\ast}(\omega \otimes \mu).
$$

\subsection{Adjoint action (algebra)}
Moving to the Lie algebra, we have another pair of adjoint actions.
The adjoint action on the Lie algebra is simply the bracket
$$
\text{ad}_{\xi} : \mathfrak{g} \to \mathfrak{g}, 
\text{ad}_{\xi}\eta = [\xi, \eta]_{\mathfrak{g}} = -[\xi,\eta]_{M}.
$$

\subsection{Dual of adjoint action (algebra)}
Fix $\omega \otimes \mu \in \mathfrak{g}^{\ast}$ and $\eta \in \mathfrak{g}$.
Then the calculation runs as follows:
\begin{align*}
\langle \text{ad}_{\xi}^{\ast}(\omega \otimes \mu), \eta \rangle &=
\langle \omega \otimes \mu, \text{ad}_{\xi} \eta \rangle \\
&= -\int_{M} \omega \cdot [\xi, \eta]_{M} \, \mu.
\end{align*}
Now we apply the following identity:
$$
\intp{[\xi,\eta]}{\omega} = \lie{\xi}{\intp{\eta}{\omega}} - 
\intp{\eta}{\lie{\xi}{\omega}}.
$$
We apply Cartan's magic formula to the first Lie derivative:
$$
\lie{\xi}{\intp{\eta}{\omega}} = 
\extd \intp{\xi}{\intp{\eta}{\omega}} + \intp{\xi}{\extd \intp{\eta}{\omega}}
$$
Note that $\intp{\eta}{\omega}$ is a function, so $\intp{\xi}{\intp{\eta}{\omega}} = 0.$
Then the divergence theorem yields
$$
\int_{M} \intp{\xi}{\extd \intp{\eta}{\omega}} \, \mu 
= -\int_{M} \intp{\eta}{\omega} \, \text{div}_{\mu} \xi \, \mu.
$$
Hence
$$
-\int_{M} \omega \cdot [\xi,\eta]_{M} \, \mu 
= \int_{M} \intp{\eta}{(\lie{\xi}{\omega} + \omega \text{div}_{\mu} \xi)} \, \mu,
$$
which means that
$$
\text{ad}_{\xi}^{\ast}(\omega \otimes \mu) = 
\left( \lie{\xi}{\omega} + \omega \text{div}_{\mu} \xi \right) \otimes \mu.
$$

\subsection{Diamond map} We start with the map $\rho_{f}$.  
Given $\xi \in \mathfrak{g}$ and $f \in V$, we define
$$
\rho_{f}(\xi) := \xi f = \lie{\xi}{f}.
$$
Then the dual map $\rho^{\ast}_{f} : V^{\ast} \to \mathfrak{g}^{\ast}$ can be
computed as follows: for $a \in V^{\ast}$,
\begin{align*}
\langle \rho^{\ast}_{f}(a), \xi \rangle &= \langle a, \rho_{f}(\xi) \rangle \\
 &= \langle a, \lie{\xi}{f} \rangle \\
 &= \int_{M} \lie{\xi}{f} \, a \\
 &= \int_{M} \extd f \cdot \xi \, a \\
 &= \langle \extd f \otimes a, \xi \rangle \\
\end{align*}
Hence $\rho^{\ast}_{f}(a) = \extd f \otimes a$.  Now we can define the diamond map,
\begin{gather*}
\diamond : V \times V^{\ast} \to \mathfrak{g}^{\ast} \\
f \diamond a = \extd f \otimes a
\end{gather*}

\subsection{Summary of Results}
Starting with
\begin{itemize}
\item $\gamma \in G = \text{Diff}(M)$,
\item $f \in V = \EuScript{C}^{\infty}(M)$,
\item $\mu \in V^{\ast} = \text{Den}(M)$,
\item $\xi, \eta \in \mathfrak{g} = \mathfrak{X}(M)$, and
\item $\theta \otimes \nu \in \mathfrak{g}^{\ast} = \Lambda^{1}(M) \otimes \text{Den}(M)$,
\end{itemize}
we summarize our work so far in the following table:
\begin{align*}
G \text{ action on } V&: \gamma f = \gamma^{\ast}(f) \\
G \text{ action on } V^{\ast}&: \gamma \mu = \gamma^{\ast}(\mu) \\
\text{Lie algebra bracket on } \mathfrak{g}&: [\xi, \eta]_{\mathfrak{g}} = -[\xi,\eta]_{M} \\
\mathfrak{g} \text{ action on } V&: \xi f = \lie{\xi}{f} \\
\mathfrak{g} \text{ action on } V^{\ast}&: \xi \mu = \lie{\xi}{\mu} \\
\text{Ad action of } G \text{ on } \mathfrak{g}&: \text{Ad}_{\gamma} \xi = \gamma_{\ast} \xi \\
\text{Ad}^{\ast} \text{ action of } G \text{ on } \mathfrak{g}^{\ast}&: \text{Ad}^{\ast}_{\gamma} (\theta \otimes \nu) = \gamma^{\ast}(\theta \otimes \nu) \\
\text{ad action of } \mathfrak{g} \text{ on } \mathfrak{g}&: \text{ad}_{\xi} \eta = [\xi, \eta]_{\mathfrak{g}} = -[\xi, \eta]_{M} \\
\text{ad}^{\ast} \text{ action of } \mathfrak{g} \text{ on } \mathfrak{g}^{\ast}&: \text{ad}^{\ast}_{\xi}(\theta \otimes \nu) = (\lie{\xi}{\theta} + \theta \, \text{div}_{\nu} \xi) \otimes \nu \\
\diamond: V \times V^{\ast} \to \mathfrak{g}^{\ast}&: f \diamond \mu = \extd f \otimes \mu
\end{align*}

\section{Unreduced Lagrangian Dynamics}
Our goal now is to present a derivation of the Euler equations for compressible
flow in material coordinates.
First we write down an expression for the unreduced Lagrangian.
Here $\eta \in \text{Diff}(M)$, $\dot{\eta} \in T_{\eta} \text{Diff}(M)$,
and $\mu_0 \in V^{\ast} = \text{Den}(M)$.  Then the Lagrangian 
$L : T \text{Diff}(M) \times V^{\ast} \to \mathbb{R}$ is
$$
L(\dot{\eta}, \mu_0) = \int_{M}
\left[ \frac{1}{2}
|| \dot{\eta} (m) ||^2 - W \Bigl(
[\eta_{\ast} \mu_0] (\eta(m)) \Bigr) \right] \mu_0.
$$
The first term in the integrand is the kinetic energy, while the second
term is the potential energy.  We assume the fluid is barotropic, meaning
that the pressure is a function only of the fluid's density.  This forces 
the potential energy to also be a function only of the fluid's density.
The density $\mu_0$ here is the material fluid density.  By this we mean that
$\mu_0$ is the density as a function of material points in the
fluid, at time $t = 0$. 

\subsection{Equations of motion}
We obtain equations of motion by applying Hamilton's principle directly
to $L$.  The calculations
that follow are almost identical to the calculations that would result from
taking $L$ and plugging it into the Euler-Lagrange equations.

First we write $\mu_0 = \rho_0 \, \mathrm{d}^n x$ for some 
$\rho_0 \in \EuScript{C}^{\infty}(M)$.  Then we have
\begin{align*}
\delta L &= \int_{M} \left[ \dot{\eta}^i \delta \dot{\eta}^i \right. \\
&-\left. W' \left( (\det \nabla \eta)^{-1} \rho_0 (m) \right)
\left( - ( \det \nabla \eta )^{-1}
(\nabla \eta)^{-T} \cdot \nabla \delta \eta 
\rho_0 \right) \right] \rho_0 \, \mathrm{d}^n x \\
 &= \int_{M} \left[ \dot{\eta}^i \delta \dot{\eta}^i \right. \\
&+\left. W' \left( (\det \nabla \eta)^{-1} \rho_0 (m) \right)
\left( (\det \nabla \eta)^{-1} 
\frac{\partial (\eta^{-1})^j}{\partial x^i}
\frac{\partial (\delta \eta)^i}{\partial X^j}
\rho_0 \right) \right] \rho_0 \, \mathrm{d}^n x,
\end{align*}
and after integrating by parts,
\begin{align*}
\delta L &= \int_{M} \left\{ - \ddot{\eta}^i \delta \eta^i \rho_0 \right. \\
 &- \left. \frac{\partial}{\partial X^j} \left[ W' \left( (\det \nabla \eta)^{-1}
 \rho_0 (m) \right) \left( (\det \nabla \eta)^{-1} 
 \frac{ \partial (\eta^{-1})^j }{ \partial x^i } \right) \rho_0^2 \right]
 \delta \eta^i \right\} \, \mathrm{d}^n x.
\end{align*}
Along solution curves, $\delta L$ must be zero for all $\delta \eta^i$, 
implying that $\eta$ must satisfy the PDE
\begin{equation}
\label{eqn:eulermaterial}
\rho_0 \ddot{\eta}^i = -\frac{\partial}{\partial X^j} \left[ W' \left(
(\det \nabla \eta)^{-1} \rho_0 (m) \right) \left( (\det \nabla \eta)^{-1}
\frac{\partial (\eta^{-1})^j}{\partial x^i} \right) \cdot \rho_0^2 \right].
\end{equation}

\subsection{Substitution of variables}
We will now try to express this PDE in the spatial variables $u$ and 
$\rho_s$.  This will enable us to compare it to standard compressible fluid
PDE.  The spatial density $\rho_s$ (a function of spatial points $x$
and time $t$) is related to the material density $\rho_0$ in the following
way:
$$
\rho_s \circ \eta = \frac{1}{\det \nabla \eta} \rho_0.
$$
Differentiating the relation
$$
u \circ \eta = \dot{\eta}
$$
gives us
$$
\ddot{\eta} = \dot{u} \circ \eta + \nabla u \cdot u \circ \eta.
$$
Substituting into (\ref{eqn:eulermaterial}), we have
\begin{align*}
&\left[ (\det \nabla \eta ) \rho_s \circ \eta \right]
(\dot{u} \circ \eta + \nabla u \cdot u \circ \eta )^i \\
&= -\frac{\partial}{\partial X^j} \left[ W' \left( \rho_s \circ \eta(m) \right)
\left( \rho_s \circ \eta \cdot \frac{\partial (\eta^{-1})^j}{\partial x^i}
(\det \nabla \eta) \rho_s \circ \eta \right) \right].
\end{align*}
We can now get rid of $\circ \eta$ everywhere to obtain
\begin{equation}
\label{eqn:eulersubst1}
(\det \nabla \eta) \rho_s (\dot{u} + \nabla u \cdot u)^i =
-\frac{\partial \eta^k}{\partial X^j} \frac{\partial}{\partial x^k}
\left[ W' (\rho_s) \rho^2_s \frac{\partial (\eta^{-1})^j}{\partial x^i}
\det \nabla \eta ] \right].
\end{equation}
We have used the following fact: 
because $x^k = \eta^k (X)$, we have, locally,
$$
\frac{\partial}{\partial X^j} = \frac{\partial x^k}{\partial X^j} 
\frac{\partial}{\partial x^k} = \frac{\partial \eta^k}{\partial X^j}
\frac{\partial}{\partial x^k}.
$$
We expand the right-hand side of (\ref{eqn:eulersubst1}) into three terms:
\begin{align*}
&- \frac{\partial \eta^k}{\partial X^j} \left\{ \left( 
\frac{\partial}{\partial x^k}  \left[ W'(\rho_s) \rho_s \right]
\frac{\partial (\eta^{-1})^j}{\partial x^i} \det \nabla \eta \rho_s \right) \right. \\
&+ \left. [W'(\rho_s) \rho_s] \left( \frac{\partial \rho_s}{\partial x^k} 
\cdot \frac{\partial (\eta^{-1})^j}{\partial x^i} \det \nabla \eta \right. \right. \\
&+ \left. \left. \rho_s \frac{\partial}{\partial x^k} \left( 
\frac{\partial (\eta^{-1})^j}{\partial x^i} \det \nabla \eta \right) \right)
\right\}.
\end{align*}
The last term is actually zero:
\begin{align*}
&\frac{\partial \eta^k}{\partial X^j} \frac{\partial}{\partial x^k}
\left( \frac{\partial (\eta^{-1})^j }{\partial x^i} (x) \cdot
\frac{1}{\det (\nabla \eta^{-1})} \right) \\
&= \left( \frac{\partial \eta^k}{\partial X^j} 
\frac{\partial^2 (\eta^{-1})^j}{\partial x^k \partial x^i} 
\frac{1}{\det (\nabla \eta^{-1})} \right)
- \frac{1}{(\det (\nabla \eta^{-1}))^2} \frac{\partial}{\partial x^i}
(\det \nabla \eta^{-1}) \\
&= \left( \frac{\partial \eta^k}{\partial X^j} 
\frac{\partial^2 (\eta^{-1})^j}{\partial x^k \partial x^i} 
\frac{1}{\det (\nabla \eta^{-1})} \right)
- \frac{1}{(\det (\nabla \eta^{-1}))^2} 
\frac{\partial \eta^{n}}{\partial X^m} \det \nabla \eta^{-1} 
\frac{\partial^2 (\eta^{-1})^m}{\partial x^i \partial x^n} \\
&= 0.
\end{align*}
This means we can simplify (\ref{eqn:eulersubst1}) all the way to the following:
$$
(\dot{u} + \nabla u \cdot u)^i = -\frac{\partial}{\partial x^i}
[W'(\rho_s) \rho_s] - W'(\rho_s) \frac{\partial \rho_s}{\partial x^i}.
$$

\subsection{Induced action of $G$ on $TG$}
Before we show that $L$ is invariant under the action of the group
$G = \text{Diff}(M)$, we revisit the action of $G$ on $TG$.  For each
$\gamma \in G$, let $\Phi_{g}: G \to G$ denote the action of $\gamma$
on $G$.  Specifically,
$$
\Phi_{\gamma}(\eta) = \eta \circ \gamma.
$$
This action induces an action on each tangent space $T_{\eta}G$.
Specifically, we have the map $T \Phi_{\gamma} : TG \to TG$ which
acts as follows.  Fix $\dot{\eta} \in T_{\eta}G$.  We can
write this element of $T_{\eta}G$ as
$\dot{\eta} = \left. \frac{d}{d \epsilon}\right|_{\epsilon=0}
\eta^{\epsilon}$ where $\eta^0 = \eta$.  Then
\begin{align*}
T_{\eta} \Phi_{\gamma} \cdot \dot{\eta} &= \left. \frac{d}{d \epsilon}
\right|_{\epsilon = 0} \Phi_{\gamma}(\eta^{\epsilon}) \\
 &= \left. \frac{d}{d \epsilon} \right|_{\epsilon = 0} \eta^{\epsilon} \circ \gamma \\
 &= \dot{\eta} \circ \gamma.
\end{align*}
Hence we obtain the action of $\gamma$ on $TG$:
$$
\gamma \cdot (\eta, \dot{\eta}) := T \Phi_{\gamma} \cdot (\eta, \dot{\eta}) = (\eta \circ \gamma, \dot{\eta} \circ \gamma).
$$
Or since $\dot{\eta} \circ \gamma \in T_{\eta} \text{Diff}(M)$, we can ignore the
effect of the action on the base point of the tangent vector and write
$$
\gamma \cdot \dot{\eta} = \dot{\eta} \circ \gamma.
$$
The group $G$ acts on $\text{Den}(M)$ via pullback, i.e. $\gamma \mu =
\gamma^{\ast} \mu$.
(Though some actions considered are actually \emph{right} actions,
we will always write the group variable on the left.)

\subsection{Proof of $G$-invariance of $L$}
We now verify that our Lagrangian $L$ is in fact $G$-invariant.
\begin{align*}
L(\gamma \cdot (\dot{\eta}, \mu))
&= L(\dot{\eta} \circ \gamma, \gamma^{\ast} \mu) \\
&= \int_{M} \left[ \frac{1}{2}
|| \dot{\eta} \circ \gamma ||^2 - W \Bigl(
\left[ (\eta \circ \gamma)_{\ast} \gamma^{\ast} \mu \right]
(\eta \circ \gamma) \Bigr)
\right] (\gamma^{\ast} \mu).
\end{align*}
Recall that
$$
(\eta \circ \gamma)_{\ast}  (\gamma^{\ast} \mu) = \eta_{\ast} \gamma_{\ast}
\gamma^{\ast} \mu = \eta_{\ast} \mu,
$$
so that
$$
L(\gamma \cdot (\dot{\eta}, \mu)) = \int_{M}
\left[ \frac{1}{2}
|| \dot{\eta} \circ \gamma (m) ||^2 - W \Bigl(
[\eta_{\ast} \mu] (\eta \circ \gamma ) (m) \Bigr) \right]
(\gamma^{\ast} \mu).
$$
Noting that $\gamma^{\ast}(f) = f \circ \gamma$, we see that
$$
L(\gamma \cdot (\dot{\eta}, \mu)) = \int_{M} \gamma^{\ast}
\left[ \frac{1}{2}
|| \dot{\eta} (m) ||^2 - W \Bigl(
[\eta_{\ast} \mu] (\eta(m)) \Bigr) \right]
(\gamma^{\ast} \mu).
$$
Now applying the change of variables theorem, we have
$$
L(\gamma \cdot (\dot{\eta}, \mu)) = \int_{M}
\left[ \frac{1}{2}
|| \dot{\eta} (m) ||^2 - W \Bigl(
[\eta_{\ast} \mu] (\eta(m)) \Bigr) \right] \mu,
$$
which is precisely $L(\dot{\eta}, \mu)$, proving $G$-invariance.

\section{Reduced Lagrangian Dynamics}
Now we can write down the reduced Lagrangian.  Here $v \in \mathfrak{X}(M)$,
and $\mu_s \in \text{Den}(M)$.
$$
l(v, \mu_s) = L(v \circ \eta, \eta^{\ast} \mu_s).
$$
Note that the
Eulerian velocity $v = \dot{\eta} \circ \eta^{-1}$, and $v$ is a
tangent vector based at the identity element of $\text{Diff}(M)$.  Also
note that $\mu_s = \eta_{\ast} \mu_0$ or $\mu_0 = \eta^{\ast} \mu_s$.
We write the right-hand side as follows:
$$
L(v \circ \eta, \eta^{\ast} \mu_s) = \int_{M} \left[
\frac{1}{2} ||v \circ \eta||^2 - W \bigl( \eta^{\ast} \mu_s \bigr) \right]
\eta^{\ast} \mu_s.
$$
Using the properties of pullback, we write
$$
L(v \circ \eta, \eta^{\ast} \mu_s) = \int_{M} \eta^{\ast} \left( \left[
\frac{1}{2} ||v||^2 - W (\mu_s) \right] \mu_s \right).
$$
Now by the change of variables theorem,
$$
l(v, \mu_s) = \int_{M} \left[ \frac{1}{2} ||v||^2 - W (\mu_s) \right] \mu_s,
$$
which is our final expression for the reduced Lagrangian
$l : \mathfrak{g} \times V^{\ast} \to \mathbb{R}$ where, of course,
$\mathfrak{g} \times V^{\ast} = \mathfrak{X}(M) \times \text{Den}(M)$.
In what follows, we will use the following notation:
$\mu_s = \rho_s \, \mathrm{d}^n x$,
where $\mathrm{d}^n x$ is the canonical $n$-form on $M$ and
$\rho_s \in \EuScript{C}^{\infty}(M)$. 

\subsection{Equations of Motion I: Variational Principle}
Our first path from the reduced Lagrangian $l$ to the equations of motion
will involve the reduced variational principle.  That is, critical points
$(v, \rho_s)$ of $l$, subject to the constraints
\begin{align*}
\delta v &= \frac{\partial w}{\partial t} + [v, w], \ \text{and} \\
\delta \rho_s &= - \frac{\partial}{\partial x^i} (\rho_s w^i),
\end{align*}
are actual trajectories of the system.  Or, in other words, if one solves
the variational equation 
$$
\delta l (v, \rho_s) \cdot (\delta v, \delta \rho_s) = 0
$$
subject to the above constraints on the variations $\delta v$ and
$\delta \rho_s$, one finds solutions to the equations of motion
that describe the evolution of the system.

\subsection{Derivation of Constraints}
Before proceeding, we confirm
that the above constraints are natural.  They arise precisely from the
relationship between the unreduced ($\eta$, $\rho_0$) and reduced
($v$, $\rho_s$) variables.  Free variations of the former correspond to
constrained variations of the latter.

To see this, we start with the relationship between the reduced and
unreduced velocity variables, $u = \dot{\eta} \circ \eta^{-1}$, and
calculate variations.  First we write
$$
u^{\epsilon} = \dot{\eta}^{\epsilon} \circ (\eta^{\epsilon})^{-1},
$$
which implies that
$$
\delta u := \left. \frac{d}{d \epsilon} \right|_{\epsilon = 0} u^{\epsilon}
= \delta \dot{\eta} \circ \eta^{-1} + \nabla \dot{\eta} \cdot
  \left. \frac{d}{d\epsilon} \right|_{\epsilon = 0} (\eta^{\epsilon})^{-1} 
$$
Now because $(\eta^{\epsilon})^{-1} \circ (\eta^{\epsilon}) = \text{id}$,
we can differentiate both sides with respect to $\epsilon$ at $\epsilon = 0$.
This yields the following formula:
$$
\left. \frac{d}{d\epsilon} \right|_{\epsilon = 0} (\eta^{\epsilon})^{-1}
 = - \nabla \eta^{-1} \cdot \delta \eta \circ \eta^{-1}.
$$
Applying this to our earlier expression for $\delta u$, we have
$$
\delta u = \delta \dot{\eta} \circ \eta^{-1} - \nabla \dot{\eta} \cdot
   \nabla \eta^{-1} \cdot \delta \eta \circ \eta^{-1}.
$$
Defining $w = \delta \eta \circ \eta^{-1}$, we find that
$$
\dot{w} = \delta \dot{\eta} \circ \eta^{-1} + \nabla \delta \eta \cdot
 \frac{\partial}{\partial t} (\eta^{-1}).
$$
We can calculate the very last term in this formula by differentiating
$\eta^{-1} \circ \eta = \text{id}$ with respect to $t$, resulting in
$$
\frac{\partial}{\partial t} \eta^{-1} = - \nabla \eta^{-1} \cdot \dot{\eta}
 \circ \eta^{-1}.
$$
Substituting into our earlier expression for $\dot{w}$, we have
\begin{align*}
\dot{w} &= \delta \dot{\eta} \circ \eta^{-1} - \nabla \delta \eta \cdot
  \nabla \eta^{-1} \cdot \dot{\eta} \circ \eta^{-1} \\
 &= \delta \dot{\eta} \circ \eta^{-1} - \nabla w \cdot u.
\end{align*}
Using what we have so far, along with the fact that 
$$\nabla u = \nabla \dot{\eta} \cdot \nabla \eta^{-1},$$
we have
$$
\delta u = \dot{w} + \nabla w \cdot u - \nabla u \cdot w = \dot{w} + [u, w].
$$
Next, using basic facts about the pullback of volume forms, we write
$$
\rho_s \circ \eta = \frac{1}{\det \nabla \eta} \rho_0.
$$
Taking variations,
$$
\delta \rho_s \circ \eta + \nabla \rho_s \cdot \delta \eta = 
 - \frac{\rho_0}{\det \nabla \eta} (\nabla \eta)^{-T} \cdot \nabla \delta \eta,
$$
where the final $\cdot$ denotes pairing of linear transformations.  That is,
$A \cdot B = \text{tr}(A^T B)$.  Now we solve for $\delta \rho_s$, obtaining
$$
\delta \rho_s = - \nabla \rho_s \cdot w - \rho_s \text{tr}(\nabla \eta^{-1} \cdot \nabla \delta \eta).
$$
In coordinates,
\begin{align*}
\text{tr}(\nabla \eta^{-1} \cdot \nabla \delta \eta) 
&= (\nabla \eta^{-1})_{ij} (\nabla \delta \eta)_{ji} \\
&= \frac{\partial (\eta^{-1})^i}{\partial x^j} \cdot
   \frac{\partial (\delta \eta)^j}{\partial X^i}  \\
&= \frac{\partial (\delta \eta \circ \eta^{-1})^j}{\partial x^j} = \text{div}(w).
\end{align*}
Therefore,
$$
\delta \rho_s = -\nabla \rho_s \cdot w - \rho_s \text{div}(w) = - \text{div}(\rho_s w).
$$

\subsection{Critical Points of $l$}
We now vary $(v, \rho_s)$ subject to the constraints which
have just been verified. 
$$
\delta l \cdot (\delta v, \delta \rho_s) = \int_{M}
\left\{ \left[ v^i \delta v^i - W'(\rho_s) \delta \rho_s \right] \rho_s +
\left[ \frac{1}{2} |v|^2 - W \right] \delta \rho_s \right\} dx
$$
Using the constraints and switching to coordinates,
\begin{align*}
 \delta l \cdot (\delta v, \delta \rho_s) &= \int_{M}
 \left\{ \left( v^i \left[ \frac{\partial w^i}{\partial t}
                         + \frac{\partial w^i}{\partial x^j} v^j
                         - \frac{\partial v^i}{\partial x^j} w^j \right]
                   + W'(\rho_s) \frac{\partial}{\partial x^i} (\rho_s w^i)
 \right\} \rho_s \right. \\
 & \ \ 
 - \left. \left( \frac{1}{2} |v^2| - W \right) \frac{\partial}{\partial x^i}
   (\rho_s w^i) \right\} \, dx.
\end{align*}
Integrating by parts,
\begin{align*}
 \delta l \cdot (\delta v, \delta \rho_s) &= \int_{M}
 \left\{ -\frac{\partial}{\partial t} (v^i \rho_s) w^i 
         -\frac{\partial}{\partial x^j} (v^i v^j \rho_s) w^i
         -v^i \frac{\partial v^i}{\partial x^j} \rho w^j \right. \\
 & \ \ \left. -\frac{\partial}{\partial x^i} [W'(\rho_s) \rho_s] \rho_s w^i
         + \frac{\partial}{\partial x^i} 
            \left[ \frac{1}{2} |v|^2 - W(\rho_s) \right] \rho_s w^i \right\}
 dx.
\end{align*} 
The above must hold for all $w$, so we have the following equation
(in what follows, we write $\rho$ for $\rho_s$ and use subscripts to
denote spatial derivatives):
\begin{align*}
& - \frac{\partial v^i}{\partial t} \rho
- v^i \frac{\partial \rho}{\partial t}
- v^i_{,j} v^j \rho
- v^i v^j_{,j} \rho 
- v^i v^j \frac{\partial \rho}{\partial x^j} \\
&- v^j v^j_{,i} \rho
- [W'(\rho) \rho]_{,i} \rho
+ [v^j v^j_{,i} - W'(\rho) \rho_{,i}] \rho = 0
\end{align*}
The third term cancels with the sixth term.  Also, the second, fourth,
and fifth terms summed together equal the left-hand side of the continuity
equation, i.e.
$$
v^i \left( \frac{\partial \rho}{\partial t} + v^j_{,j} \rho +
 v^j \frac{\partial \rho}{\partial x^j} \right) = 0.
$$
After eliminating these terms, we cancel a factor of $\rho$ and obtain
$$
\frac{\partial v^i}{\partial t} + v^i_{,j} v^j = 
- \left[ W'(\rho) \rho_{,i} + (W'(\rho) \rho)_{,i} \right].
$$
Not surprisingly, this is precisely the same equation we got by taking 
free variations of the unreduced Lagrangian and \emph{then} changing from 
material to spatial coordinates.

\subsection{Equations of Motion II: Euler-Poincar\'{e} Equation}
We can also derive the equations of motion from the reduced Lagrangian
by making use of the Euler-Poincar\'{e} equation from semidirect product
reduction theory:
$$
\frac{\partial}{\partial t} \frac{\delta l}{\delta v} 
= - \text{ad}^{\ast}_{v} \frac{\delta l}{\delta v}
  + \frac{\delta l}{\delta \mu_s} \diamond \mu_s.
$$

\subsection{Derivatives of $l$}
In order to apply the Euler-Poincar\'{e} equations of motion, we must 
first calculate the variational derivatives of $l(v, \mu_s).$  First
\begin{align*}
\biggl \langle \frac{\delta l}{\delta v}, \xi \biggr \rangle &= 
\left. \frac{d}{d \epsilon}
\right|_{\epsilon = 0} l(v_{\epsilon}, \mu_s) \text{ where } \left.
\frac{d}{d \epsilon} \right|_{\epsilon = 0} v_{\epsilon} = \xi \\
 &= \left. \frac{d}{d \epsilon} \right|_{\epsilon = 0} \int_{M} \frac{1}{2}
|| v_{\epsilon} ||^2 \, \mu_s \\
 &= \int_{M} \frac{1}{2} \left. \frac{d}{d \epsilon} \right|_{\epsilon = 0}
    || v_{\epsilon} ||^2 \, \mu_s \\
 &= \int_{M} \langle v, \xi \rangle_{|| \cdot ||} \, \mu_s \\
 &= \langle v^{\flat} \otimes \mu_s, \xi \rangle
\end{align*}
where the $\flat$ operation is pointwise with respect to 
$\langle \cdot, \cdot \rangle_{|| \cdot ||}$, i.e.
$$
v^{\flat}(\xi) = \langle v, \xi \rangle_{|| \cdot ||}.
$$
Hence 
$$
\frac{\delta l}{\delta v} = v^{\flat} \otimes \mu_s.
$$
Note that $W$ only depends on the $\rho_s$ ``part'' of $\mu_s$, not on the
canonical volume form $\mathrm{d}^n x$ part.  Therefore,
$$
W(\mu_s) = W(\rho_s \, \mathrm{d}^n x) = \widehat{W} (\rho_s).
$$
Here $W : \text{Den}(M) \to \EuScript{C}^{\infty}(M)$.  
Using $\widehat{W}$, we can write an alternate expression for 
$\mathbf{D} W : \text{Den}(M) \to \mathbf{L}(\text{Den}(M), \EuScript{C}^{\infty}(M))$ as follows.  Let $\mu = \rho \, \mathrm{d}^n x$ and $\mu' = \rho' \, \mathrm{d}^n x$.
Then
\begin{align*}
\mathbf{D} W (\mu) \cdot \mu' &= \left. \frac{d}{d \epsilon} \right|_{\epsilon = 0}
W(\mu + \epsilon \mu') \\
&= \left. \frac{d}{d \epsilon} \right|_{\epsilon = 0} \widehat{W} (\rho + \epsilon \rho') \\
&= \frac{\partial \widehat{W}}{\partial \rho} (\rho) \cdot \rho' = \widehat{W}'(\rho) \rho'.
\end{align*}
Furthermore, for densities $\mu$, $\mu_1$, and $\mu_2$, we have
\begin{align*}
\langle \mathbf{D} W(\mu) \cdot \mu_1, \mu_2 \rangle &=
\biggl \langle \widehat{W}'(\rho) \rho_1, \mu_2 \biggr \rangle \\
&= \int_{M} \widehat{W}'(\rho) \rho_1 \mu_2 \\
&= \int_{M} \widehat{W}'(\rho) \rho_1 \rho_2 \, \mathrm{d}^n x \\
&= \int_{M} \widehat{W}'(\rho) \rho_2 \mu_1 \\
&= \biggl \langle \widehat{W}'(\rho) \rho_2, \mu_1 \biggr \rangle \\
&= \langle \mathbf{D} W(\mu) \cdot \mu_2, \mu_1 \rangle
\end{align*}
Armed with these facts, we proceed:
\begin{align*}
\biggl \langle \frac{\delta l}{\delta \mu_s}, \alpha \biggr \rangle &= \left. 
\frac{d}{d \epsilon} \right|_{\epsilon = 0} l(v, \mu_s^{\epsilon})
\text{ where } \left. \frac{d}{d \epsilon} \right|_{\epsilon  = 0} 
\mu_s^{\epsilon} = \alpha \\
&= \int_{M} \left. \frac{d}{d \epsilon} \right|_{\epsilon = 0}
\left[ \frac{1}{2} || v ||^2 - W(\mu_s^{\epsilon}) \right] \mu_s^{\epsilon} \\
&= \int_{M} \left[ \frac{1}{2} || v||^2  - W(\mu) \right] \alpha +
 \mu_s \left[ - \mathbf{D} W(\mu_s) \cdot \alpha \right] \\
&= \biggl \langle \frac{1}{2} ||v||^2 - W(\mu_s), \alpha \biggr \rangle - 
   \langle \mathbf{D} W(\mu_s) \cdot \alpha, \mu_s \rangle \\
&= \biggl \langle \frac{1}{2} ||v||^2 - W(\mu_s) 
   - \mathbf{D} W(\mu_s) \cdot \mu_s, \alpha \biggr \rangle,
\end{align*}
so we must have
$$
\frac{\delta l}{\delta \mu_s} = \frac{1}{2} ||v||^2 - W(\mu_s) - \widehat{W}'(\rho_s) \cdot \rho_s.
$$
\subsection{Plugging into Euler-Poincar\'{e}}
In what follows, we write $\mu$ to mean $\mu_s$.  In an earlier calculation,
we established that for $\xi \in \mathfrak{g}$ and 
$\theta \otimes \nu \in \mathfrak{g}^{\ast}$, 
$$
\text{ad}^{\ast}_{\xi} (\theta \otimes \nu) = (\lie{\xi}{\theta}
 + \theta \text{div}_{\nu} \xi) \otimes \nu.
$$
Applying this in our case yields
$$
\frac{\partial}{\partial t} (v^{\flat} \otimes \mu) 
 = -(\lie{v}{v^{\flat}} + v^{\flat} \text{div}_{\mu} v) \otimes \mu
   + \frac{\delta l}{\delta \mu} \diamond \mu.
$$
Now because $\text{div}_{\mu} v$ is a scalar function, the two objects
$- v^{\flat} \text{div}_{\mu} v \otimes \mu$ and
$- v^{\flat} \otimes \text{div}_{\mu} v \mu$ are equivalent.  Thus we
can pick off the advection equation
$$
v^{\flat} \otimes \frac{\partial \mu}{\partial t} = 
- v^{\flat} \otimes (\text{div}_{\mu} v) \mu,
$$
which simplifies to
$$
\frac{\partial \mu}{\partial t} = - (\text{div}_{\mu} v) \mu = -\lie{v}{\mu}.
$$
Then we can simplify our earlier equation:
$$
\frac{\partial v^{\flat}}{\partial t} \otimes \mu 
 = - \lie{v}{v^{\flat}} \otimes \mu + \extd \left( 
   \frac{\delta l}{\delta \mu} \right) \otimes \mu.
$$
We have made use of our previous computation for the diamond map.  Now 
we apply Cartan's magic formula and substitute for the remaining derivative:
$$
\frac{\partial v^\flat}{\partial t} + \extd \intp{v}{v^{\flat}} + \intp{v}{\extd v^{\flat}}
= \frac{1}{2} \extd \intp{v}{v^{\flat}} - \extd W(\mu) - \extd (\mathbf{D} W(\mu) \cdot \mu).
$$
A coordinate computation verifies that, in fact,
$$
\frac{1}{2} \extd \intp{v}{v^{\flat}} + \intp{v}{\extd v^{\flat}} = [(v \cdot \nabla) v]^{\flat}.
$$
Taking sharps of both sides, we end up with
$$
\frac{\partial v}{\partial t} + (v \cdot \nabla) v =
- \left[ \extd W(\mu) + \extd (\mathbf{D} W(\mu) \cdot \mu) \right]^{\sharp},
$$
which agrees with our previous results perfectly.

\section{Hamiltonian Dynamics} We will attempt a Hamiltonian derivation
of the following system:
\begin{equation}
\label{eqn:onedimeuler}
\begin{split}
\partial_t \rho + \left( \rho u \right)_{x} &= 0 \\
\partial_t \left( \rho u \right) + \left( \rho u^2 + p \right)_{x} &= 0.
\end{split}
\end{equation}
These are the Euler equations for a compressible fluid on the line.  The equations are also known as the \emph{continuity} and \emph{momentum} equations, respectively, of one-dimensional gas dynamics.

\subsection{Setup}  We know we can write the one-dimensional Lagrangian as a function of a velocity $u$ and a density $\mu = \rho \, dx$:
\begin{equation}
\label{eqn:lagr}
l(u,\rho) = \frac{1}{2} \int \rho \| u \|^2 \, dx + \int \rho W(\rho) \, dx
\end{equation}
A simple computation reveals that $\delta l / \delta u = \rho u$, which is therefore the correct momentum variable to use in the Hamiltonian formulation.  Let us set $\mathbf{m} = \rho u$ and write $H$ as a function of $\mathbf{m}$ and $\rho$:
\begin{equation}
\label{eqn:ham1}
H(\mathbf{m}, \rho) = \frac{1}{2} \int \frac{1}{\rho} \| \mathbf{m} \|^2 \, dx
+ \int \rho W(\rho) \, dx.
\end{equation}
Here $H$ is a function on the semidirect product Lie algebra $\mathfrak{g} \times V$.
Typically we view $H$ as a function on the dual of a Lie algebra, but in this case it does not matter.  The reason is that we can identify $\mathbf{m}$ with the one-form density
$$
\mathbf{\bar{m}} = v^{\flat} \otimes \rho \, dx
$$
and we can also identify $\rho$ with $\rho \, dx$.  With these identifications, we can write $H(\mathbf{\bar{m}}, \rho \, dx)$ and its expression will be identical to (\ref{eqn:ham1}).  Now we can apply the semidirect product reduction theorem, obtaining the Lie-Poisson equations on the dual algebra $(\mathfrak{g} \times V)^\ast$.  (Note to reader: I will make this more concrete and explicit in the next draft.  For now I will focus on the computations.)

\subsection{Brackets} The semidirect product Poisson bracket is given in our case by
\begin{equation}
\label{eqn:bracket}
\{ F, G \} (\mathbf{m}, \rho) = \int \mathbf{m} \cdot \left[\frac{\delta F}{\delta \mathbf{m}}, \frac{\delta G}{\delta \mathbf{m}} \right] \, dx - \int \rho \left(
 \lie{\delta F / \delta \mathbf{m}}{\frac{\delta G}{\delta \rho}}
 - \lie{\delta G / \delta \mathbf{m}}{\frac{\delta F}{\delta \rho}} \right) \, dx.
\end{equation}
The bracket under the first integral sign is the Lie algebra bracket on $\mathfrak{g}$.  It is a standard result that this bracket is \emph{minus} the ordinary Jacobi-Lie bracket of vector fields on $M$.  Therefore in one dimension we have
\begin{align}
\label{eqn:bracket1}
\{F, G \} (\mathbf{m}, \rho) = &\int \mathbf{m} \left( \frac{\delta G}{\delta \mathbf{m}} \left( \frac{\delta F}{\delta \mathbf{m}} \right)_{x} - \frac{\delta F}{\delta \mathbf{m}} \left( \frac{\delta G}{\delta \mathbf{m}} \right)_{x} \right) \, dx \\
 &+ \int \rho \left( \frac{\delta G}{\delta \mathbf{m}} \left( \frac{\delta F}{\delta \rho} \right)_{x}
  - \frac{\delta F}{\delta \mathbf{m}} \left( \frac{\delta G}{\delta \rho} \right)_{x} \right) \, dx
\end{align}  

\subsection{Derivatives of $H$} We shall use (\ref{eqn:ham1}) and compute variational derivatives.  First,
\begin{align*}
\left \langle \frac{\delta H}{\delta \mathbf{m}}, \delta \mathbf{m} \right \rangle &=
 \left. \frac{d}{d \epsilon} \right|_{\epsilon = 0} H(\mathbf{m} + \epsilon \, \delta \mathbf{m},\rho) \\
&= \left. \frac{d}{d\epsilon} \right|_{\epsilon = 0} \left[ \frac{1}{2} \int \frac{1}{\rho}
 ( \mathbf{m} + \epsilon \, \delta \mathbf{m})^2 \, dx \right] \\
&= \int \frac{1}{\rho} \mathbf{m} \, \delta \mathbf{m} \, dx
\end{align*}
Hence $\delta H / \delta \mathbf{m} = (1/\rho) \mathbf{m}$.  Next,
\begin{align*}
\left \langle \frac{\delta H}{\delta \rho}, \delta \rho \right \rangle &=
 \left. \frac{d}{d \epsilon} \right|_{\epsilon = 0} H(\mathbf{m}, \rho + \epsilon \, \delta \rho) \\
  &= \left. \frac{d}{d\epsilon} \right|_{\epsilon = 0} \frac{1}{2} \int
  \frac{1}{\rho + \epsilon \, \delta \rho} \mathbf{m}^2 \, dx + \int (\rho + \epsilon
  \, \delta \rho) W(\rho + \epsilon \, \delta \rho) \, dx \\
  &= \frac{1}{2} \int -\frac{1}{\rho^2} \, \delta \rho \, \mathbf{m}^2 \, dx +
   \int (W(\rho) + \rho W'(\rho)) \, \delta \rho \, dx 
\end{align*}
Hence $\delta H / \delta \rho = - (1/2) \mathbf{m}^2 / \rho^2 + W(\rho) + \rho W'(\rho)$.

\subsection{Computing the Equations of Motion}  We start with $F(\mathbf{m},\rho) = \int \mathbf{m} \, dx$.  In this case, $\delta F / \delta \rho = 0$ and
$$
\left \langle \frac{\delta F}{\delta \mathbf{m}}, \delta \mathbf{m} \right \rangle = \int \delta \mathbf{m} \, dx
$$
so $\delta F / \delta \mathbf{m} = 1$.  (There must be a more technically accurate way of writing this, but for now this will have to do.)  We are now in a position to apply the general evolution equation
$$
\dot{F}(\mathbf{m}, \rho) = \left\{ F, H \right \} (\mathbf{m}, \rho).
$$
We have
\begin{equation}
\label{eqn:momderiv1}
\begin{split}
\int \partial_t \mathbf{m} \, dx = &\int \mathbf{m} \frac{\mathbf{m}}{\rho} (1)_{x} \, dx \\
& -\int \mathbf{m} \, (1) \left( \frac{\mathbf{m}}{\rho} \right)_{x} \, dx \\
& -\int \rho \, (1) \left( -\frac{1}{2} \frac{\mathbf{m}^2}{\rho^2} + W(\rho) + \rho W'(\rho) \right)_{x} \, dx
\end{split}
\end{equation}
In the above expression, we have hung on to the symbol $(1)$ everywhere to remind ourselves that this is not actually the number one but instead an element of the space $\mathfrak{g}^\ast$, i.e. a one-form density on $M$.  When applied to vector fields, the map $(1)$ acts like the number one.  Otherwise we make no assumptions regarding its spatial derivatives and charge ahead by integrating the first term in (\ref{eqn:momderiv1}) by parts to obtain
$$
\int \mathbf{m} \frac{\mathbf{m}}{\rho} (1)_{x} \, dx = - \int \left( \frac{\mathbf{m}^2}{\rho} \right)_{x} (1) \, dx.
$$
We get out from underneath the integral sign and also substitute $\mathbf{m} = \rho u$
\begin{equation}
\label{eqn:momeqn1}
\partial_t \left( \rho u \right) = - \left( \rho u^2 \right)_{x} - \rho u u_{x} + \rho u u_{x} - p_{x}.
\end{equation}
Here $p = \rho^2 W'(\rho)$ is the pressure.  A quick computation reveals that
$$
p_x = \rho \, \left( W + \rho W' \right)_{x}.
$$
Thus our final result for the momentum equation is
\begin{equation}
\label{eqn:momeqn2}
\partial_t \left( \rho u \right) + \left( \rho u^2 + p \right)_{x} = 0.
\end{equation}
Next we take $F(\mathbf{m}, \rho) = \int \rho \, dx$.  In this case, $\delta F / \delta \mathbf{m} = 0$ and just like before,
$$
\left \langle \frac{\delta F}{\delta \rho}, \delta \rho \right \rangle = \int \delta \rho \, dx,
$$
so we shall write $\delta F / \delta \rho = 1$, where we mean essentially the same thing by ``1'' as we did earlier.  Now the only term from the bracket $\{F, H\}$ that survives is the one involving $\delta F / \delta \rho$.  Hence
$$
\int \partial_t \rho \, dx = \int \rho \frac{\delta H}{\delta \mathbf{m}} (1)_{x} \, dx 
 = \int \mathbf{m} (1)_{x} \, dx 
 = - \int \mathbf{m}_{x} (1) \, dx.
$$
Now substituting $\mathbf{m} = \rho u$ and pulling things out of the integral sign we have
\begin{equation}
\label{eqn:conteqn}
\partial_t \rho + \left( \rho u \right)_{x} = 0.
\end{equation}
Equations (\ref{eqn:conteqn}) and (\ref{eqn:momeqn2}) give us the system (\ref{eqn:onedimeuler}).

\end{document}